\documentclass[twocolumn,amsmath,amssymb,prd]{revtex4-2}

\def\al{\alpha}
\def\be{\beta}

\def\de{\delta}
\def\ep{\epsilon}

\def\et{\eta}
\def\th{\theta}

\def\la{\lambda}

\def\si{\sigma}

\def\ph{\phi}

\def\ch{\chi}

\def\om{\omega}

\def\De{\Delta}

\def\La{\Lambda}

\def\Om{\Omega}

\def\fr#1#2{{{#1}\over{#2}}}
\def\frac#1#2{{\textstyle{{#1}\over{#2}}}}

\def\Re{\hbox{Re}\,}
\def\Im{\hbox{Im}\,}

\def\lsim{\mathrel{\rlap{\lower4pt\hbox{\hskip1pt$\sim$}}
    \raise1pt\hbox{$<$}}}
\def\gsim{\mathrel{\rlap{\lower4pt\hbox{\hskip1pt$\sim$}}
    \raise1pt\hbox{$>$}}}

\def\etal{{\it et al.}}

\def\vev#1{\langle {#1}\rangle}

\def\ket#1{|{#1}\rangle}

\def\eff{{\rm eff}}

\def\sqr#1#2{{\vcenter{\vbox{\hrule height.#2pt
         \hbox{\vrule width.#2pt height#1pt \kern#1pt
         \vrule width.#2pt}
         \hrule height.#2pt}}}}

\newcommand{\beq}{\begin{equation}}
\newcommand{\eeq}{\end{equation}}
\newcommand{\bea}{\begin{eqnarray}}
\newcommand{\eea}{\end{eqnarray}}
\newcommand{\rf}[1]{(\ref{#1})}

\newcommand{\bM}{\begin{pmatrix}}
\newcommand{\eM}{\end{pmatrix}}

\def\nn{\nonumber}

\def\f{w}

\def\mbf#1{\boldsymbol #1}

\def\syjm#1#2{{}_{#1}Y_{#2}}

\def\V{\mathcal V}

\def\T{\mathcal T}

\def\K{\mathcal K}

\def\pvec{\mbf p}

\def\sivec{\mbf\si}

\def\bevec{\mbf\be}

\def\pmag{|\pvec|}

\def\punit{\hat p}

\def\epunit{\hat\ep}
\def\thunit{\hat\th}
\def\phunit{\hat\ph}

\def\phat{\mbf\punit}

\def\ephat{\mbf\epunit}
\def\thhat{\mbf\thunit}
\def\phhat{\mbf\phunit}

\def\gt{\widetilde g}
\def\Ht{\widetilde H}

\def\Htf#1#2#3{{{\tilde{H}}_{#1}}\,\hspace{-1 pt}^{(#2)#3}_\eff}
\def\gtf#1#2#3{{{\tilde{g}}_{#1}}\,\hspace{-1 pt}^{(#2)#3}_\eff}

\def\X{X}
\def\Y{Y}

\def\Xhat{\widehat\X}
\def\Yhat{\widehat\Y}

\def\nr{{\rm NR}}

\def\nrtemplate#1#2#3{#1^{\nr#3}_{#2}}

\def\cs133{\rm Cs}

\def\Vnrf#1#2{\nrtemplate{{\V_{#1}}}{#2}{}}
\def\TzBnrf#1#2{\nrtemplate{{\T_{#1}}}{#2}{(0B)}}

\def\ToBnrf#1#2{\nrtemplate{{\T_{#1}}}{#2}{(1B)}}

\def\anrf#1#2{\nrtemplate{{a_{#1}}}{#2}{}}
\def\cnrf#1#2{\nrtemplate{{c_{#1}}}{#2}{}}

\def\sVnrf#1#2{\nrtemplate{{\V_{#1}}}{#2}{,{\rm Sun}}}
\def\sanrf#1#2{\nrtemplate{{a_{#1}}}{#2}{,{\rm Sun}}}
\def\scnrf#1#2{\nrtemplate{{c_{#1}}}{#2}{,{\rm Sun}}}
\def\sgzBnrf#1#2{\nrtemplate{{g_{#1}}}{#2}{(0B), {\rm Sun}}}
\def\sgoBnrf#1#2{\nrtemplate{{g_{#1}}}{#2}{(1B),{\rm Sun}}}
\def\sHzBnrf#1#2{\nrtemplate{{H_{#1}}}{#2}{(0B),{\rm Sun}}}
\def\sHoBnrf#1#2{\nrtemplate{{H_{#1}}}{#2}{(1B),\rm Sun}}
\def\sTzBnrf#1#2{\nrtemplate{{\T_{#1}}}{#2}{(0B),{\rm Sun}}}
\def\sToBnrf#1#2{\nrtemplate{{\T_{#1}}}{#2}{(1B),\rm Sun}}

\def\widecheck#1{\hskip#1pt\huge$\check{}$}
\def\bighacek#1#2{\vbox{\ialign{##\crcr\widecheck#2\crcr
  \noalign{\kern-9.5pt\nointerlineskip}
   $\hfil\displaystyle{#1}\hfil$\crcr}}}

\def\k{k}

\def\chM{\vartheta}

\def \j{j}

\def\Ht{\widetilde{H}}
\def\gt{\widetilde{g}}
\def\Va#1#2#3{V_{#1}^{(#2)#3}}

\def\Jj#1{{J_#1}}

\def\TL{T_L}

\begin{document}

\title{
Prospects for testing CPT and Lorentz symmetry with deuterium ground-state Zeeman-hyperfine transitions
} 

\author{ Arnaldo J.\ Vargas$^1$}

\affiliation{
$^1$Laboratory of Theoretical Physics, Department of Physics, University of Puerto Rico, R\'io Piedras, Puerto Rico 00936\\
}
\begin{abstract}
This work presents a model for testing Lorentz and CPT symmetry through sidereal-variation studies of the hyperfine-Zeeman deuterium ground-state transition frequencies. It represents an advancement over previous models by using a well-established  deuteron wave-function parametrization to calculate contributions from nucleon Lorentz-violating operators toward the Lorentz-violating frequency shift. Furthermore, this work extends the analysis beyond the zeroth-boost order previously considered. This study centers on deuterium's potential for testing Lorentz-violating nonminimal terms. Specifically, it compares the prospects of an ongoing deuterium experiment with the current best limits on nonminimal coefficients. The conclusion drawn is that the deuterium experiment holds the potential to enhance and establish first-time limits on nonminimal proton, neutron, and electron SME coefficients, marking it as a valuable experiment in the current worldwide systematic search for Lorentz and CPT violation.
\end{abstract}

\maketitle

\section{Introduction}

Lorentz symmetry, a global symmetry of the standard model of particle physics and a local one of the theory of general relativity, plays a pivotal role in the current physics paradigm. Therefore, the pursuit of Lorentz violation as a low-energy signal for theories beyond the standard model and general relativity is unsurprising \cite{ksp}. Another compelling motivation for searching for Lorentz violation is its close connection to CPT symmetry due to the well-known result that CPT violation in a realistic field theory implies Lorentz violation \cite{owg,ck}.

A framework for the systematic search for Lorentz violation called the Standard-Model Extension (SME) \cite{ck} was proposed and successfully motivated many experimental studies in the last decades \cite{tables}. Its first iteration, the minimal SME, was constrained to Lorentz-violating operators with mass dimensions $d\leq4$. Since then, the SME has been extended, on multiple occasions, to incorporate operators with higher mass dimensions, called nonminimal Lorentz-violating operators \cite{nonmingrav, km12, km13, km09}. 

The constraints on Lorentz violation obtained from these experiments are limits on the so-called coefficients for Lorentz violation, also known as SME coefficients \cite{tables}. These coefficients are parameters that quantify the contribution of Lorentz-violating operators to the experimental observables, with each coefficient associated with a unique operator \cite{ck}. Hence, the discovery of a non-zero coefficient implies evidence of a breaking in Lorentz symmetry. The coefficients associated with the operators with mass dimensions $d\leq4$ are called minimal coefficients, and the ones linked to operators with $d>4$ are nonminimal. The coefficients are assumed to be different for operators corresponding to different types of particles, leading to categories of coefficients like electron coefficients or photon coefficients.

Since the inception of the SME, there has been a recognition that the high precision attainable in atomic spectroscopy experiments made them favorable for detecting Lorentz violation \cite{kl99,bkr}. Consequently, several experimental groups in this field designed experiments that imposed constraints on minimal SME coefficients\cite{maser, xema, xema2, Xe09, Xe14, cs06, cs17, NeRb, ma13, hu01, ah18}. 

 In recent years, there has been an effort to study the prospects of using atomic spectroscopy experiments to detect Lorentz-violating signals from nonminimal operators \cite{gkv14, kv15, kv18}. These studies reported limits on nonminimal SME coefficients by reinterpreting the results from experiments designed in the context of the minimal SME framework. For instance and of relevance for the discussion, a reassessment \cite{kv15} of the results obtained by a time-variation study with a hydrogen maser \cite{maser} derived constraints on nonminimal SME electron and proton coefficients that contributed to the Lorentz-violating signals constrained in the experiment.  

These publications also delved into determining systems better suited for imposing bounds on nonminimal coefficients than those proposed in the minimal SME context. A difference between minimal Lorentz-violating contributions to the atomic spectrum and the nonminimal ones is that some nonminimal coefficients are proportional to higher powers of the momentum of the fermions within the atom than the minimal coefficients. This observation led to the recognition that using atoms with higher momentum constituents could enhance sensitivity to these coefficients \cite{gkv14, kv15, kv18}.

Consequently, hydrogen might not be the optimal choice for limiting the size of some proton coefficients as every nonexotic atom, except for hydrogen, possesses a multi-nucleon nucleus that enhances the sensitivity to these coefficients. The rationale behind this lies in the fact that the internal nuclear motion of the proton adds to its overall momentum, which, as previously mentioned, is a desirable attribute. A simplified deuterium model and order-of-magnitude calculations demonstrated that conducting a similar experiment to the hydrogen-maser one with deuterium could substantially improve previously reported coefficient limits \cite{kv15}.

The heightened sensitivity to the proton coefficients applies to any atom with a multi-nucleon nucleus, not solely deuterium, as highlighted in a subsequent publication \cite{kv18} to the one that presented the deuterium model.  However, considering these heavier atoms presents a challenge, compared to the hydrogen case, as understanding the internal nuclear state becomes crucial to derive Lorentz-violating corrections to their spectrum.  

The commonly used nuclear model within the SME framework \cite{kl99, kv18} is the Schmidt model \cite{s37, bw52} due to its comparative simplicity. The model assumes a shell structure for the nucleus, where pairs of nucleons of the same type form states with zero total angular momentum. Nevertheless, this model's drawback is that it suggests many atomic systems are sensitive to either neutron or proton coefficients but not both, contrary to expectations where both types of coefficients are anticipated to contribute, albeit with the contribution from one type being suppressed compared to the other. For example, within the context of the minimal SME, applying more advanced techniques to obtain the energy shift unveiled contributions from coefficients associated with both nucleons contrary to the prediction from Schmidt model \cite{st15,fl16,cs17}. Moreover, while the model serves to approximate angular momentum expectation values, it is indifferent to the contributing nucleon's momentum magnitude and requires it to be estimated by other means. Despite its limitations, the simplicity of the Schmidt model enabled the imposition of stringent limits on numerous nonminimal SME coefficients \cite{kv18}. 

In this context, the Stefan Meyer Institute (SMI) in Vienna has initiated a deuterium beam spectroscopy experiment \cite{SMI1}, currently underway at the Laboratoire Aim\'e  Cotton (LAC) in Paris \cite{SMIpriv}. This experiment is motivated by the aforementioned deuterium model presented in \cite{kv15}. However, as a model designed to demonstrate the potential advantage of deuterium over hydrogen, it remained underdeveloped to work as a functional model for imposing bounds on SME coefficients.

This work aims to develop a model for the Lorentz-violating shift to the hyperfine-Zeeman transition frequencies of deuterium's ground state, in the presence of a weak magnetic field, with the expectation to be used by the SMI/LAC experiment and potentially by future similar experiments to establish bounds on nonminimal SME coefficients. One of deuterium's advantages lies in its multiple well-established deuteron ground-state wave-function parametrizations. For this study, we will employ the parametrization based on the Paris nucleon-nucleon potential \cite{parisnn1, parisnn2}. This study marks the first application of a nuclear model beyond the Schmidt model in the nonminimal SME context. Moreover, deuterium's sensitivity to neutron and proton coefficients positions it as an excellent candidate for establishing new and enhanced bounds on nonminimal SME coefficients.

This paper contains three additional sections besides the introduction and summary. The subsequent section, Sec. \ref{seclab}, outlines the derivation of the laboratory-frame Lorentz-violating frequency shift for the Zeeman-hyperfine transitions within the ground state of deuterium. Following is Sec. \ref{secsun}, which presents the transformation of the frequency shift into the canonical Sun-centered frame, the main result of this work. The discussion of the prospects for the SMI/LAC experiment in establishing new or enhanced limits on SME coefficients is the topic of Sec. \ref{secprosp}.  Finally, throughout this work, natural units with $\hbar=c=1$ are employed.

\section{Lorentz-violating frequency shift}
\label{seclab}

This section, comprising four subsections, derives the Lorentz-violating frequency shift for the Zeeman-hyperfine transitions of the ground state of deuterium. The first two subsections introduce the unperturbed deuterium states and the Lorentz-violating perturbation considered in this work. Following is a calculation of the relevant Lorentz-violating correction to the deuterium's spectrum in the presence of a weak magnetic field. The section concludes by deriving an expression for the Lorentz-violating frequency shift for the pertinent transitions. 

\subsection{Unperturbed quantum states}

In this subsection, we present a simplified nonrelativistic three-fermion system model for the deuterium atom in its zero-momentum inertial frame. We use this simplified model to obtain the unperturbed states that will be used in the perturbative calculation.

The deuterium atom is a bound system composed of a proton, a neutron, and an electron. The electron orbiting motion around the nucleus is approximately nonrelativistic. Likewise, we can apply a similar nonrelativistic approximation to describe the proton and neutron motion inside the deuteron. While this approximation may not provide the same level of precision for describing the motion of nucleons as it does for the electron, it remains a valid approach as the three-momentum of the nucleons is smaller than their masses in natural units.

The unperturbed Hamiltonian considered in this work has the form  

\beq
H_{\rm D} \approx
\fr{\mbf{p}_e^2}{2m_e}+\fr{\mbf{p}_p^2}{2m_{\rm p}}+\fr{\mbf{p}_n^2}{2m_n}
+ V + U,
\label{hD}
\eeq
where $\mbf p_\f$ represents the three-momentum of each particle, with  $e$ for the electron,  $p$ for the proton, and $n$ for the neutron.  The mass of the corresponding fermion is denoted by  $m_\f$ with $\f\in\{ e,p,n\}$. The Hamiltonian includes two interaction potentials.  The potential $V$ represents the electromagnetic interaction between the proton and electron, and $U$ is the nuclear interaction between the proton and neutron.

Another approximation employed in this work is the treatment of the deuteron as being nearly at rest in the zero-momentum frame of the deuterium atom. This assumption is reasonable because the deuteron is significantly more massive than the electron. Under this premise, the momenta of the proton and neutron are opposite, $\pvec_{p}\simeq-\pvec_{n}$, in the deuterium zero-momentum frame. To simplify the presentation, we introduce a notation where $\pvec\simeq\pvec_{p}$, which implies $\pvec_{n}\simeq-\pvec$.

The potential $V(r_{ep})$ in the Hamiltonian \eqref{hD} is a function of $r_{ep}$, which represents the distance between the electron and proton. Due to the large size of the deuterium radius compared to the deuteron one, we can approximate $r_{ep}\simeq r_{ed}$, where $r_{ed}$ is the distance between the electron and the center of mass of the deuteron. The potential $U(\mathbf{r}_{pn})$, on the other hand, depends on the position of the proton relative to the neutron. 
 
Building upon the earlier discussion, we can decompose the Hamiltonian \eqref{hD} into two parts as follows
\beq
H_D \approx H_e+H_d.
\eeq
The Hamiltonian $H_e$ represents the single-particle Hamiltonian governing the behavior of the electron in the deuterium zero-momentum frame. It can be expressed as
\beq
H_e \approx \fr{\mbf{p}_e^2}{2 m_e}+ V(r_{ed}),
\label{eH}
\eeq
where $r_{ed}$ is the distance between the electron and the center of mass of the deuteron.
Conversely, $H_d$ encapsulates the Hamiltonian describing the behavior of the deuteron in its rest frame
\beq
H_d\approx\fr{\mbf{p}^2}{2 \mu_d} + U(\mbf{r_{pn}}).
\label{dH}
\eeq
Here,  $\mu_d=m_p m_n /(m_p +m_n)$ is the reduced mass in the proton-neutron two-body system.  

We denote the single-particle states of the electron within the deuterium atom as $\ket{nJLm}$, where the quantum numbers $(J, m)$ label the total electronic angular momentum, $L$ represents the orbital angular momentum, and $n$ signifies the principal quantum number. They are the energy states of the Hamiltonian \eqref{eH}, and the ones used in this work are the well-established nonrelativistic solutions for an electron in the presence of a coulomb potential.  

The ground state of the Hamiltonian \eqref{dH} is denoted as $\ket{S_d m_d}$, where $(S_d, m_d)$ are the quantum numbers associated with the spin of the deuteron. We can express the quantum state for the deuterium atom with total atomic angular momentum quantum numbers $(F, m_F)$ as 
\beq
\ket{n F J L m_F} =
\sum_{m ,m_d} 
\vev{J m S_d m_d|F m_F}
\ket{nJLm}\ket{S_d m_d}
\eeq
\\
where $\vev{j_1 m_1 j_2 m_2|j_3 m_3}$ represents the Clebsch-Gordan coefficients.

Our primary focus in this study lies in the deuterium $1S_{1/2}^F$ states. For this reason, it proves advantageous to introduce a more concise notation by omitting certain quantum numbers, specifically $n=1$, $J=1/2$, $L=0$, and $S_d=1$. Under this abbreviated notation, we represent the deuterium state $\ket{1, F ,1/2 ,0 ,m_F}$ as $\ket{ F m_F}$, the single-electron state $\ket{1 ,1/2 ,0, m}$ as $\ket{m}$, and the deuteron state $\ket{1 m_d}$ as $\ket{ m_d}$. Consequently, $\ket{F m_F}$ takes the form
\beq
\ket{F m_F} =
\sum_{m ,m_d} 
\vev{1/2, m, 1, m_d|F m_F}
\ket{m}\ket{m_d}.
\label{Fstates}
\eeq

It is important to note that the state $\ket{m_d}\equiv\ket{1 m_d}$ does not constitute an eigenstate of the total orbital angular momentum of the nucleons within the deuteron due to the non-central nature of the potential $U(\mathbf{r}_{pn})$. The deuteron ground state is a superposition of the $^3S_{1}$ and $^3D_{1}$ internal nuclear states.  Hence, we can approximate the momentum wave function of the deuteron ground state as follows
\bea
\vev{\pvec|m_d}&=&
u_{0}(p) Y_{00}(\phat) \ch_{m_d}\nn \\
&&+ u_{2}(p)\sum_{qm}
\vev{1q2m|1m_d} Y_{2m}(\phat) \ch_{q}.
\eea
In this equation, $\vev{j_1 m_1 j_2 m_2|j_3 m_3}$ denotes the Clebsch-Gordan coefficients, the summation over $q$ and $m$ encompasses all allowed values, $u_l(p)$ denotes functions that depend solely on $p=\pmag$, $Y_{jm}(\phat)$ denotes the spherical harmonics evaluated in the direction of $\phat$, and $\chi_m$, with $m\in\{-1,0,1\}$, represents the spin-triplet state derived from the proton and neutron spin states. 

In this study, we will employ the parametrization of the deuteron wave function based on the Paris nucleon-nucleon potential in momentum space, as detailed in \cite{parisnn1, parisnn2}. In this parametrization, the functions $u_{0}(p)$ and $u_{2}(p)$ are given by
\bea
u_0(p)&=&\sqrt{\fr{2}{\pi}} \sum_{j=1}^n \fr {C_j}{p^2+m_j^2 },\nn\\
u_2(p)&=&\sqrt{\fr{2}{\pi}} \sum_{j=1}^n \fr {D_j}{p^2+m_j^2 }.
\label{Pnnup}
\eea
The coefficients $C_j$, $D_j$, and $m_j$ are provided in \cite{parisnn2}. 

It's worth noting that our  main goal in this study is to estimate the magnitude of the coefficients for Lorentz violation, rather than conducting precise calculations of the energy shift. Thus, any of the commonly used analytical parametrizations for the deuteron wave function could have been suitable for this work. However, we have opted for this specific parametrization due to its convenience for performing some of the integrals needed for this study.

\subsection{Lorentz-violating perturbation}

In principle, the Lorentz-violating perturbation can receive contributions from both the free propagation of fermions within the atom and corrections to their interactions. Previous studies \cite{kv15,kv18} have established that the dominant contribution to the Lorentz-violating energy shift in deuterium results from the perturbation affecting the free propagation of the fermions within the atom. Contributions arising from corrections to the interactions will be treated as higher-order corrections to the energy shift and, as such, will not be considered in this work.

Therefore, in this work, we will treat the Lorentz-violating perturbation to the deuterium atom as the sum of three components
\beq
\de h=\de h_{e}^\nr+\de h_{p}^\nr+\de h_{ n}^\nr,
\label{hexpa}
\eeq
where $\de h_w^\nr$ represents the single-particle Lorentz-violating perturbation for the particle with flavor $w$. 

The derivation of a comprehensive Lorentz-violating perturbation for a freely-propagating  nonrelativistic Dirac fermion is provided in \cite{km13}. Certain Lorentz-violating operators contributing to this perturbation, as demonstrated in previous works \cite{gkv14, kv15, kv18}, do not contribute to the atomic energy shift at first order in perturbation theory. Therefore, the effective single-particle Lorentz-violating perturbation considered in this work, consistent with previous works \cite{gkv14, kv15, kv18}, can be expressed as follows
 \bea
\de h_\f^\nr
&=& -\sum_{k \j m} \pmag^k 
\syjm{0}{\j m}(\phat) 
\left(\Vnrf{\f}{\k \j m}+\sivec\cdot\ephat_r\TzBnrf{\f}{k \j m}\right)\nn\\
&&+\sum_{k \j m} \sivec\cdot\left(\syjm{+ 1}{\j m}(\phat) \ephat_--\syjm{- 1}{\j m}(\phat) \ephat_+\right)
  \ToBnrf{\f}{k \j m}\nn\\
\label{nr}
\eea
with $\sivec=(\si^1,\si^2,\si^3)$ standing for the Pauli vector composed by the Pauli matrices $\si_i$.  

The summation index $k$ takes on the values 0, 2, and 4, while the index $j$ ranges from 0 to 5, and the index $m$ is over the range $-j\leq m\leq j$. The unit vectors in the equation are defined as $\ephat_r = \phat$ and $\ephat_\pm = (\thhat \pm i\phhat)/\sqrt{2}$. These definitions are in terms of the unit vectors $\thhat$ and $\phhat$ associated with the polar angle $\th$ and azimuthal angle $\ph$ in momentum space. These angles can be specified by the relation $\phat = (\sin\th\cos\ph,\sin\th\sin\ph,\cos\th)$.

The functions $\syjm{s}{\j m}(\phat)$ correspond to the spin-weighted spherical harmonics with a spin weight of $s$. The conventional spherical harmonics $Y_{\j m}(\th,\ph)$ can be understood as spin-weighted harmonics with $s=0$ and can be expressed as $\syjm{0}{\j m}(\th, \ph)$.  Additional information about the definitions and useful properties of spin-weighted spherical harmonics can be found in Appendix A of \cite{km09}.

The coefficients $\Vnrf{\f}{k \j m}$ and ${\T_\f}^{\nr(qB)}_{k\j m}$, where $q$ can take values of $0$ or $1$, represent nonrelativistic spherical coefficients for Lorentz violation as defined in \cite{km13}. Each nonrelativistic coefficient can be expressed as a linear combination of the standard SME coefficients, the ones in the SME Lagrange density, appropriately weighted by powers of $m_\f$. The explicit expressions for these combinations can be found in equations  (111) and (112) of \cite{km13}. Table IV of the same reference provides information about the permissible ranges of values for the indices $k$, $\j$, $m$, and the number of independent components for each coefficient. It's worth noting that, in this work, we adhere to the convention established in \cite{kv15}, employing the subscript index $k$ instead of $n$, which is the convention used in \cite{km13}.

 It is common practice to further decompose the components of the perturbation Hamiltonian based on their CPT handedness. Specifically, each nonrelativistic spherical coefficient can be divided into two parts, each characterized by the CPT handedness of the corresponding operator. This decomposition can be expressed as follows
\bea
\Vnrf{\f}{k\j m} &=&
\cnrf{\f}{k\j m} - \anrf{\f}{k\j m},
\nonumber \\
{\T_\f}^{\nr(qP)}_{k\j m} &=&
{g_\f}^{\nr(qP)}_{k\j m} - {H_\f}^{\nr(qP)}_{k\j m},
\label{cpt}
\eea 
where $a$- and $g$-type coefficients are associated with CPT-odd operators, while the $c$- and $H$-type coefficients are related to CPT-even ones. This notation aligns with the standard assignments in the minimal SME \cite{ck}.

\subsection{General Form of the Lorentz-Violating Energy Shift for States in $1S_{1/2}^F$ with $F\leq3/2$}

As mentioned in the introduction, this work focuses on the Zeeman-hyperfine transitions within the ground state of deuterium. We can obtain the relevant energy shifts from the matrix elements of the Lorentz-violating perturbation within the subspace spanned by the states $\ket{Fm_F}$ as given in \eqref{Fstates} with $F\leq3/2$. Furthermore, the assumed presence of a weak magnetic field breaks the degeneracy associated with the orientation of the total deuterium angular momentum ${\mbf F}$ allowing the use of the methods of nondegenerate perturbation theory. Consequently, we can determine the leading-order Lorentz-violating energy shift by evaluating the expectation values of $\de h$, as defined in \eqref{hexpa}, for the states $\ket{Fm_F}$.

It is convenient to decompose the expectation value of the perturbation, the Lorentz-violating energy shift,  as follows
\beq
\de \ep=\vev{\de h}=\vev{\de h_e^\nr}+\vev{\de h_d},
\eeq
where $\de h_d$ represents the perturbation to the deuteron, given by $\de h_d = \de h_p^\nr + \de h_n^\nr$.

The expectation value of the perturbation $\de h_d$ for the states $\ket{Fm_F}$ takes the following form
\bea
\vev{\de h_d}&=&-\sum_{w= p,n }\sum_k\left(\fr {\vev{\pmag^k}} {\sqrt{4\pi}}\Vnrf{\f}{k00}+\vev{\syjm{0}{20}(\phat)\pmag^k } \Vnrf{\f}{k20}\right)\nn\\
&&+2\sum_{w=p,n }\sum_k\vev{\left(\syjm{+ 1}{10}(\phat)  \pmag^k\sivec\cdot\ephat_-\right) }\ToBnrf{\f}{k10}\nn\\
&&-\sum_{w=p,n}\sum_k\vev{\syjm{0}{10}(\phat) \pmag^k \sivec\cdot\ephat_r }\TzBnrf{\f}{k10},
\label{expn}
\eea
where the summation over $k$ takes values of 0, 2, and 4. Additionally, it should be noted that the coefficient $\Vnrf{\f}{020}$ does not exist, as per the properties of the nonrelativistic coefficients discussed in \cite{km13}. Therefore, we can set it equal to zero in the expression \eqref{expn}. Furthermore, any terms in \eqref{nr} that are absent from \eqref{expn} have vanishing expectation values for $F\leq 3/2$.  

The expectation values in \eqref{expn} have the form
\bea
\vev{\pmag^k}&=&\vev{u_0|\pmag^k|u_0}+\vev{u_2|\pmag^k|u_2},\nn\\
\vev{ \syjm{0}{20}(\phat)\pmag^k}&=&\fr{2F-1}{(10-8 m_F^2)\sqrt{5\pi}}\vev{\pmag^k}_{0E},\nn\\
\vev{ \syjm{0}{10}(\phat)\pmag^k\sivec\cdot\ephat_r  }&=&\fr{2m_F}{3(2F+1)\sqrt{3\pi}} \vev{\pmag^k}_{0B},\nn\\
\vev{\syjm{+ 1}{10}(\phat)  \pmag^k \sivec\cdot\ephat_- }&=&\fr{m_F}{3(2F+1)\sqrt{3\pi}}\vev{\pmag^k}_{1B}
\label{pexp}
\eea
with $F\leq 3/2$. The values of $\vev{u_l|\pmag^k|u_m}$ are calculated from
\beq
\vev{u_l|\pmag^k|u_m}=\int_0^\infty  u_l^*(p) u_m(p) p^{k+2} dp,
\eeq
where the functions $u_l(p)$ are described in  \eqref{Pnnup}. 

The coefficients $\vev{\pmag^k}_{0E}$, $\vev{\pmag^k}_{0B}$, and $\vev{\pmag^k}_{1B}$ in \eqref{pexp} are linear combinations of the expectation values $\vev{u_l|\pmag^k|u_m}$, and they have the following form
\bea
\vev{\pmag^k}_{0E}& = &\vev{u_2|\pmag^k|u_2}-\sqrt{8}\Re{\vev{u_0|\pmag^k|u_2}}, \nn\\
\vev{\pmag^k}_{0B}&=&2\vev{u_0|\pmag^k|u_0}+\vev{u_2|\pmag^k|u_2}\nn\\
&&+\sqrt{8}\Re{\vev{u_0|\pmag^k|u_2}}\nn\\
\vev{\pmag^k}_{1B}&=&-4\vev{u_0|\pmag^k|u_0}+4\vev{u_2|\pmag^k|u_2}\nn\\
&&+\sqrt{8}\Re{\vev{u_0|\pmag^k|u_2}},
\eea
where $\Re[x]$ denotes the real part of $x$.  

Numerical values for these coefficients are outlined in Table \ref{npexp}. The first column of the table indicates the corresponding expectation values or coefficients. The subsequent columns contain the numerical values for each instance of $k$, specifically for  $0$, $2$, and $4$. Entries marked as N/A indicate instances where the value was not calculated because it is not needed.  
\renewcommand\arraystretch{1.5}
\begin{table}
\caption{
Numerical values for the coefficients in Eq. \eqref{pexp}.} 
\setlength{\tabcolsep}{6pt}
\begin{tabular}{cccc}
\hline
\hline
Coefficient & k=0 & k=2 (${\rm GeV}^2$) & k=4 (${\rm GeV}^4$)	\\
\hline
$\vev{\pmag^k}$           & 1.0       &  $1.8\times 10^{-2}$        &   $4.1\times 10^{-3}$\\
$\vev{\pmag^k}_{0E}$ & N/A      &  $7.8\times 10^{-3} $        &   $-1.4\times 10^{-3}$\\
$\vev{\pmag^k}_{0B}$ & $1.7$   &  $ 2.8\times 10^{-2} $        &  $9.7\times  10^{-3} $\\
$\vev{\pmag^k}_{1B}$ & $-3.8$  &  $ -1.2\times 10^{-2}$      &   $3.9\times 10^{-3} $\\
\hline\hline
\end{tabular}
\label{npexp}
\end{table}

The expectation value of the electron component of the perturbation  $\de h$ is
\bea
\vev{\de h_{\rm e}^\nr}&=&\fr{2(1-F)m_F}{3\sqrt{3\pi}}\sum_k \vev{|\pvec_{e}|^k}\left(\TzBnrf{ e}{k10}+2\ToBnrf{\rm e}{k10}\right)\nn\\
&&-\sum_k\fr {\vev{|\pvec_{e}|^k}} {\sqrt{4\pi}}\Vnrf{e}{k00}
\label{expe}
\eea
Once more, the summation over $k$ is performed for values of 0, 2, and 4. The expectation value of $|\pvec_{e}|^k$ is 
\beq
\vev{|\pvec_{e}|^0}=1,\; \vev{|\pvec_{e}|^2}=(\al m_r)^2,\; \vev{|\pvec_{e}|^4}=5(\al m_r)^4,
\label{pmage}
\eeq
where $\al$ represents the fine-structure constant, while $m_r$ is the reduced mass of the atom, approximately equal to the mass of the electron $m_{\rm e}$. The expressions for the energy shift in \eqref{expe} are valid for $F\leq 3/2$.  

\subsection{Lorentz-violating frequency shift for the transition $F=3/2\rightarrow F=1/2$ in a weak magnetic field}

Using the Lorentz-violating energy shift obtained in the previous subsection, we are able to compute the Lorentz-violating frequency shift for the Zeeman-transitions $(F=1/2, m_F')\rightarrow (F=3/2,m_F)$. This frequency shift takes the form
\bea
2 \pi \de \nu &=&-\fr{1}{(5-4 m_F^2)\sqrt{5\pi}}\sum_{w={\rm p,n}} \sum_{k}\vev{\pmag^k}_{0E} \Vnrf{\f}{k20}\nn \\
&&+\fr{2m'_F-m_F}{6\sqrt{3\pi}}\sum_{w={\rm p,n}} \sum_{k}\vev{\pmag^k}_{0B}\TzBnrf{\f}{k10}\nn\\
&&-\fr{2m'_F-m_F}{6\sqrt{3\pi}}\sum_{w={\rm p,n}} \sum_{k}\vev{\pmag^k}_{1B}\ToBnrf{\f}{k10}\nn\\
&&-\fr{m_F+m'_F}{3\sqrt{3\pi}}\sum_k \vev{|\pvec_{e}|^k}\left(\TzBnrf{\rm e}{k10}+2\ToBnrf{\rm e}{k10}\right).\nn\\
\label{LVfshift}
\eea
We remind the reader that the summation over $k$ is limited to the values 0, 2, and 4. Additionally, the coefficient $ \Vnrf{\f}{020}$ should be regarded as zero, as this particular combination of indices is not permitted, as explained in \cite{km13}.

We can disregard the contribution from the electron coefficients to the frequency shift. This is due to the fact that hydrogen maser experiments exhibit significantly greater sensitivity to these coefficients than what one would reasonably anticipate from a deuterium experiment. The rationale behind this observation was initially addressed in \cite{kv15}, and we will elaborate on it further below. 

Before delving into the discussion, let us reproduce the result for the Lorentz-violating frequency shift obtained in \cite{kv15} for the hydrogen hyperfine-Zeeman transitions of the ground state. The frequency shift is given by 
\bea
2\pi\de\nu &=&
-\fr{\De m_F}{2\sqrt{3\pi}}
\sum_k\sum_{\f =e,p}
\vev{|\pvec_e|^k}\left(\TzBnrf{\f}{k10}+2\ToBnrf{\f}{k10}\right),\nn\\
\label{hyd}
\eea
where  $\vev{|\pvec_e|^k}$ is given by \eqref{pmage}, with the reduced mass $m_r$ being the one for hydrogen, which can also be taken as the mass of the electron.  Here, $\De m_F$ represents the change in the quantum number $m_F$ during the transition.  

Keep in mind that the sensitivity to the electron coefficients in \eqref{hyd} and \eqref{LVfshift} is proportional to the ratio $\de \nu_{\rm exp}/\vev{|\pvec_e|^k}$ of the experimental constraint on the frequency shift $\de \nu_{\rm exp}$ relative to the corresponding expectation value $\vev{|\pvec_e|^k}$. The expectation values $\vev{|\pvec_e|^k}$ are nearly identical for hydrogen and deuterium. Therefore, the experiment using either hydrogen or deuterium that imposes the most stringent constraint on the corresponding frequency shift will achieve the highest sensitivity to the electron coefficient. A previous study with a hydrogen maser \cite{maser} constrained the Lorentz-violating frequency shift to around 0.1 mHZ. Assuming that a deuterium experiment could obtain constraints on the order of 1 Hz, it follows that the sensitivity of the hydrogen experiment to the electron coefficients surpasses that achievable by the deuterium experiment by around four orders of magnitude. As we will see in the next section, this argument only permits us to disregard the electron coefficients at zeroth-boost order because the hydrogen-maser experiment did not consider linear-boost order effects, leaving the door open to imposing bounds on electron coefficients using deuterium's experiments.

The situation is different for the proton coefficients contributing to the hydrogen and deuterium frequency shifts in \eqref{hyd} and \eqref{LVfshift}, respectively. In the case of hydrogen, the sensitivity remains proportional to the ratio $\de \nu_{\rm exp}/\vev{|\pvec_e|^k}$, where $\vev{|\pvec_e|^2}\sim 10^{-11} {\rm GeV}^2$ and $\vev{|\pvec_e|^4}\sim 10^{-23} {\rm GeV}^4$. However, for deuterium, the sensitivity is represented by either $\de \nu_{\rm exp}/ \vev{\pmag^k}_{0B}$ or $\de \nu_{\rm exp}/ \vev{\pmag^k}_{1B}$. Using the values in Table \ref{npexp}, we observe that the sensitivity to the proton coefficients $\TzBnrf{p}{k10}$ and $\ToBnrf{p}{k10}$ is enhanced due to the difference in the proton’s momentum between the two atoms. Despite the heightened sensitivity to the frequency shift $\de \nu_{\rm exp}$ in the hydrogen experiment, a hypothetical deuterium experiment exhibits significantly higher sensitivity to the proton coefficients.

Assuming the aforementioned values for the constraints on the frequency shift, the deuterium experiment demonstrates approximately 5 orders of magnitude greater sensitivity to the proton coefficients with $k=2$ and approximately 15 orders of magnitude greater for those with $k=4$ than the hydrogen one. This highlights the primary motivation for pursuing Lorentz-violation studies with deuterium.

\section{Frequency shift in the Sun-centered frame}
\label{secsun}
This section presents the derivation of the expression for the frequency shift in the Sun-centered frame. The methodology used has been thoroughly discussed in previous works \cite{kv15,kv18}. The first subsection introduces the frequency shift in the Sun-centered frame at zeroth-boost order, while the subsequent subsection presents the result at the first-boost order.

\subsection{Frequency shift at zeroth-boost order}
The SME coefficients are assumed to remain constant and uniform across all inertial reference frames \cite{ck}. However, in non-inertial frames, these coefficients can exhibit time-dependent variations. Moreover, they are frame-dependent as they transform as tensor components under observer transformations. Consequently, all experiments must report limits on SME coefficients within the same inertial reference frame to permit any comparison of results across different experiments. The Sun-centered celestial-equatorial frame is used in the literature as the standard frame for this purpose \cite{sunframe}. Although the rest frame of the Sun does not precisely qualify as an inertial reference frame, any possible time variation of the coefficients in this frame is insignificant compared to the time scale between the different experiments.

In the Sun-centered frame, the origin is at the Sun's location during the 2000 vernal equinox, and the time coordinate $T$ is the Sun's proper time. The spatial Cartesian coordinates $X^J\equiv (X, Y, Z)$, in this frame, have the $Z$ axis aligned along the Earth's rotation axis, the $X$ axis pointing from the Earth to the Sun at $T=0$, and the $Y$ axis completing a right-handed coordinate system.

The matrix for the observer Lorentz transformation from the laboratory frame to the Sun-centered frame has the form
\beq
\La^\mu_{\ \nu}({\mbf\th},\bevec)=
\mathcal{R}^\mu_{\ \al}({\mbf \th})\mathcal{B}^\al_{\ \nu}(\bevec),
\label{lortr}
\eeq
where $\mathcal{R}^\mu_{\ \nu}({\mbf \th})$ represents a rotation matrix, and $\mathcal{B}^\mu_{\ \nu}(\bevec)$ a boost matrix. Here, the boost parameter $\bevec$ is the velocity of the laboratory frame within the Sun-centered frame, and the rotation parameter $\mbf \th$ specifies the orientation between the laboratory frame and the boosted one.

The magnitude of $\bevec$ is relatively small compared to the speed of light, with a value of $\be\simeq10^{-4}$ in natural units. We can simplify the expression for $\La^\mu_{\ \nu}$ by expanding it as a power series of $\be$ and truncating it at a certain power of $\be$. Truncating the power series at the zeroth-boost order in $\be$ simplifies the Lorentz transformation to a pure rotation, where the boost matrix in Eq. \eqref{lortr} is replaced with the identity matrix. In this subsection, we exclusively focus on the expression for the frequency shift \rf{LVfshift} in the Sun-centered frame at zeroth-boost order. The contributions at linear-boost order will be examined in the following subsection.

The coefficients in \eqref{LVfshift} are defined within a laboratory frame with its z-axis aligned to the applied magnetic field. Relating these coefficients to the ones within the Sun-centered frame requires the rotation from the Earh-based laboratory frame to the Sun-centered frame. The properties of the nonrelativistic coefficients aid in this transformation as they transform as components of dual spherical tensors under observer rotations.

The reader should bear in mind that the nonrelativistic coefficient indices relevant for the coefficient transformation are $j$ and $m$, where the index $j$ specifies the rank and $m$ the component of the spherical tensor associated with the coefficient. The meaning of these indices explains why only coefficients with $m=0$ contribute to the frequency shift \eqref{LVfshift} as they correspond to the components of a spherical tensor projected along the direction of the applied magnetic field within the laboratory frame.

The transformation rule for a generic laboratory frame coefficient ${\K_\f}_{kj0}^{\rm NR}$, with $m=0$ as they appear in \eqref{LVfshift}, in terms of the corresponding coefficient in the Sun-centered frame ${\K_\f}_{kjm}^{\rm NR, Sun}$ is given by
\beq
{\K_\f}_{kj0}^{\rm NR} =
\sum_{m} e^{i m\om_\oplus \TL}
d^{j}_{0m}(-\chM)
{\K_\f}^{\rm NR,Sun}_{kjm}.
\label{ltos}
\eeq
Here, $\chM$ denotes the angle between the applied magnetic field and the Earth's rotation axis, while the quantities $d^{j}_{mm'}(-\chM)$ refer to the small Wigner matrices, as provided in (136) of Ref. \cite{km09}, evaluated at $-\chM$. Keep in mind that the non-relativistic coefficients with $m\ne0$, which are components of a spherical tensor, are complex numbers, and they obey the condition
\beq
{(\K_\f}_{kjm}^{\rm NR})^*=(-1)^m{\K_\f}_{kj(-m)}^{\rm NR},
\eeq
where $*$ denotes complex conjugation.

The angle $\chM$ can be determined based on the local orientation of the magnetic field and the experiment's colatitude $\chi$. The corresponding expression is
\beq
\cos{\chM}= \cos{\th_l} \cos{\chi}+\sin{\th_l} \sin{\chi} \sin{\ph_l},
\label{chM}
\eeq
where $\ph_l$ represents the local cardinal direction of the magnetic field measured from the local East in a counterclockwise orientation. For instance, $\ph_l=0$ corresponds to the local East, and $\ph_l=\pi/2$ corresponds to the local North. The angle $\th_l$ measures the orientation relative to the local vertical direction, where $\th_l=0$ indicates that the magnetic field points towards the zenith and $\th_l=\pi/2$ implies that it is horizontal.

The SMI/LAC experiment was located at the Laboratoire Aimé Cotton, located at the University of Paris-Saclay, with a colatitude of approximately $\chi \simeq 41.3^\circ$ \cite{SMIpriv}. The applied magnetic field was horizontal, with $\th_l \simeq \pi/2$ and $\ph_l \simeq 84^\circ$. Using \rf{chM}, the angle between the magnetic field and the Earth's rotation axis is approximately $\chM \simeq 49^\circ$.

The transformation \rf{ltos} reveals the time variation, decomposed into harmonics of the Earth's sidereal frequency $\om_\oplus\simeq 2\pi/(23{\rm ~h} ~56{\rm ~min})$, of the laboratory-frame coefficients. The local sidereal time $\TL$ serves as a practical measure of the Earth's local sidereal time, with $T_L=0$ chosen as the time when the magnetic field is perpendicular to $\hat{Y}$ in the Sun-centered frame with a nonnegative $X$ component. This particular choice results in the relatively straightforward expression \rf{ltos}. It is important to note that $\TL$ is an offset from the time $T$ in the Sun-centered frame by an amount that depends on the orientation of the magnetic field. Since we will be introducing another local sidereal time later in this work, we will designate $T_L$ as the local sidereal time adjusted to the orientation of the applied magnetic field.

Applying the transformation \rf{ltos} to the coefficients contributing to the frequency shift \rf{LVfshift}, we find that
\bea
\TzBnrf{\f}{k10}&=&\cos{\chM}\sTzBnrf{\f}{k10}\nn\\
&&-\sqrt{2} \sin{\chM}\,\Re{\left[\sTzBnrf{\f}{k11}\right]}\cos{\om_\oplus \TL}\nn\\
&&+\sqrt{2} \sin{\chM}\,\Im{\left[\sTzBnrf{\f}{k11}\right]}\sin{\om_\oplus \TL},\nn\\
\Vnrf{\f}{k20}&=&-\sqrt{\dfrac{3}{2}} \sin{2 \chM}\Re{\left[\sVnrf{\f}{k11}\right]}\cos{\om_\oplus \TL}\nn\\
&&+\sqrt{\dfrac{3}{2}} \sin{2 \chM}\Im{\left[\sVnrf{\f}{k11}\right]}\sin{\om_\oplus \TL}\nn\\
&&+\sqrt{\dfrac{3}{2}} \sin^2{\chM}\Re{\left[\sVnrf{\f}{k22}\right]}\cos{2\om_\oplus \TL}\nn\\
&&-\sqrt{\dfrac{3}{2}} \sin^2{\chM}\Im{\left[\sVnrf{\f}{k22}\right]}\sin{2\om_\oplus \TL}\nn\\
&&+\dfrac{1}{4}(1+3\cos{2\chM})\sVnrf{\f}{k20},
\label{trans}
\eea
where $\ToBnrf{\f}{k10}$ obeys the same equation than $\TzBnrf{\f}{k10}$.  Here, $\Re{[{\K_\f}_{kj0}^{\rm NR}]}$ is the real part of the coefficient and  $\Im{[{\K_\f}_{kj0}^{\rm NR}]}$ its imaginary part. 

The general form of the frequency shift in the Sun-centered frame at zeroth-boost order is obtained by replacing the coefficients in the laboratory frame with the coefficients in the Sun-centered frame using the relations in \eqref{trans}. The final result has the form
\bea
{2\pi}\de \nu^{(0)}&=&   A_{0}^{(0)}+A_{c}^{(0)}\cos{(\om_\oplus \TL)}+A_{s}^{(0)}\sin{(\om_\oplus \TL)}\nn\\
&&+ A_{c2}^{(0)}\cos{(2\om_\oplus \TL)}+A_{s2}^{(0)}\sin{(2\om_\oplus \TL)}.
\label{sunshift0}
\eea
Here, the superscript $(0)$ is used to indicate that this expression solely includes the zeroth-boost order contribution to the frequency shift.
The amplitudes in the expression are linear combinations of the nonrelativistic coefficients in the Sun-centered frame. The amplitude for the constant term is
\bea
A_{0}^{(0)}&=&\cos{\chM}\fr{2m'_F-m_F}{6\sqrt{3\pi}}\sum_{\f,k}\vev{\pmag^k}_{0B}\,\sTzBnrf{\f}{k10} \nn\\
&&-\cos{\chM}\fr{2m'_F-m_F}{6\sqrt{3\pi}}\sum_{\f,k}\vev{\pmag^k}_{1B}\,\sToBnrf{\f}{k10}\nn\\
&&-\fr{(1+3\cos{2\chM})}{4(5-4 m_F^2)\sqrt{5\pi}}\sum_{\f,k}\vev{\pmag^k}_{0E}\, \sVnrf{\f}{k20}.
\label{A0}
\eea
The amplitudes for the first harmonic of the sidereal frequency are given by
\bea
A_{c}^{(0)}&=&-\sin{\chM}\fr{2m'_F-m_F}{3\sqrt{6\pi}}\sum_{\f,k}\vev{\pmag^k}_{0B}\,\Re{\left[\sTzBnrf{\f}{k11}\right]}\nn\\
&&+\sin{\chM}\fr{2m'_F-m_F}{3\sqrt{6\pi}}\sum_{\f,k}\vev{\pmag^k}_{1B}\,\Re{\left[\sToBnrf{\f}{k11}\right]}\nn\\
&&+\fr{\sin{2 \chM}}{(5-4 m_F^2)}\sqrt{\dfrac{3}{10\pi}}\sum_{\f,k}\vev{\pmag^k}_{0E}\, \Re{\left[\sVnrf{\f}{k21}\right]} ,\nn\\
A_{s}^{(0)}&=&\sin{\chM}\fr{2m'_F-m_F}{3\sqrt{6\pi}}\sum_{\f,k}\vev{\pmag^k}_{0B}\,\Im{\left[\sTzBnrf{\f}{k11}\right]}\nn\\
&&-\sin{\chM}\fr{2m'_F-m_F}{3\sqrt{6\pi}}\sum_{\f,k}\vev{\pmag^k}_{1B}\,\Im{\left[\sToBnrf{\f}{k11}\right]}\nn\\
&&-\fr{\sin{2 \chM}}{(5-4 m_F^2)}\sqrt{\dfrac{3}{10\pi}}\sum_{\f,k}\vev{\pmag^k}_{0E}\,\, \Im{\left[\sVnrf{\f}{k21}\right]} , \nn\\
\label{A1}
\eea
and the amplitudes for the second harmonic of the sidereal frequency are
\bea
A_{c2}^{(0)}&=&\fr{-\sin^2{\chM}}{(5-4 m_F^2)}\sqrt{\dfrac{3}{10\pi}} \sum_{\f,k}\vev{\pmag^k}_{0E} \Re{\left[\sVnrf{\f}{k22}\right]},\nn\\
A_{s2}^{(0)}&=&\fr{\sin^2{\chM}}{(5-4 m_F^2)}\sqrt{\dfrac{3}{10\pi}} \sum_{\f,k}\vev{\pmag^k}_{0E} \Im{\left[\sVnrf{\f}{k22}\right]}.\nn\\ 
\label{A2}
\eea

Note that the contribution from the electron coefficients was disregarded for the reasons discussed in the previous section. Only the proton and neutron coefficients were considered, thus the summation over the index $\f$ in the expressions for the amplitudes can only take the values $\f=p$ or $\f=n$. Similarly, the summation over $k$ is limited to the values 0, 2, and 4.

In these expressions, the coefficient $\sVnrf{\f}{02m}$ must be taken as zero for any value of $m$ for the reasons explained earlier. The angle $\chM$ represents the angle between the applied magnetic field and the rotation axis of the Earth. The quantum number $m’_F$ corresponds to the $F=1/2$ state involved in the transition, and $m_F$ represents the quantum number for the $F=3/2$ state. The numerical values for the relevant coefficients $\vev{\pmag^k}_{0E}$, $\vev{\pmag^k}_{0B}$, and $\vev{\pmag^k}_{1B}$ are provided in Table \ref{npexp}.  

The expression \eqref{sunshift0} for the frequency shift implies that a signal for Lorentz violation is a sidereal variation of the resonant frequency, arising from the Earth's rotation relative to a fixed inertial reference frame. This sidereal variation may include contributions up to the second harmonic of the sidereal frequency. Additionally, another signal for Lorentz violation is a dependency of the resonant frequency with the orientation of the applied magnetic field in relation to the Earth's rotation axis.

\subsection{Frequency shift at linear-boost order}

In this subsection, we derive the frequency shift in the Sun-centered frame at the linear-boost order. We begin by truncating the expansion of the observer Lorentz transformation matrix $\La^\mu_{\ \nu},({\mbf\th},\bevec)$, in the small parameter $\be \simeq 10^{-4}$, up to linear order instead of zeroth order. The resulting transformation matrix is
\beq
\La^{0}{}_{T}=1,  
\hskip 8pt
\La^{0}{}_{J}=-\bevec^J, 
\hskip 8pt
\La^{j}{}_{T}=-\mathcal{R}^{j}{}_{J}\bevec^{J},
\hskip 8pt
\La^{j}{}_{J}=\mathcal{R}^{j}{}_{J},
\label{LTlinear}
\eeq
where the lower-case and upper-case indices represent spatial Cartesian coordinates in the laboratory frame and in the Sun-centered frame, respectively.  

The Earth-based laboratory-frame boost velocity $\bevec$ within the Sun-centered frame is approximately given by
\beq
\bevec\simeq\bevec_\oplus +\bevec_L,
\label{boostpar}
\eeq
where $\bevec_\oplus$ is the velocity of the Earth relative to the Sun and $\bevec_L$ is the velocity of the laboratory relative to Earth's center of mass.  

Assuming the Earth's orbit is circular, we find that the expression for $\bevec_\oplus$ in the Sun-centered frame is 
\beq
\bevec_\oplus
=
\be_\oplus\sin{\Om_\oplus T} ~\widehat{X}
-\be_\oplus \cos{\Om_\oplus T}
( \cos\et~\widehat{Y} + \sin\et~\widehat{Z} ).  
\label{vorb}
\eeq
Here,  Earth's orbital speed is denoted by $\be_\oplus \simeq 10^{-4}$ and its orbital angular frequency by $\Om_\oplus\simeq 2\pi/(365.26 \text{ d})$.  The angle $\et\simeq23.4^\circ$ is the one between the $XY$ plane and the Earth's orbital plane.  The time $T$ is the time coordinate in the Sun-centered frame. 

The expression for 
 $\bevec_L$ obtained by taking the Earth as a sphere is
\beq
\bevec_L
=
r_\oplus \om_\oplus \sin{\ch}\left(-\sin{\om_\oplus T_\oplus}~\Xhat
+\cos{\om_\oplus T_\oplus}~\Yhat\right) ,
\label{vrot}
\eeq
where $\ch$ and $\om_\oplus$ represent, once again, the colatitude of the experiment and the sidereal frequency, respectively.  The magnitude of $\be_L$ can be estimated by using that $r_\oplus \om_\oplus\simeq 10^{-6}$, where $r_\oplus$ is Earth's radius.  This implies that $\be_L$ is two orders of magnitude smaller than $\be_\oplus$.   

The time $T_\oplus$ represents another offset of the time coordinate $T$, serving as a measurement of the local sidereal time. This definition of the sidereal time simplifies the expression \rf{vrot} by absorbing any possible phase. The zero value of $T_\oplus$ can be defined as any time when the boost velocity $\bevec_L$ is parallel to the $Y$-axis in the Sun-centered frame. Equivalently, it is the time when the local East direction, the direction of $\bevec_L$, is parallel to $\hat{Y}$. This definition differs from that of $T_L$, which is zero when the magnetic field is perpendicular to the $Y$-axis with a nonnegative $X$ component.  The relation between both local sidereal times is determined by the equations 
\bea
\cos{\left[\left(T_\oplus-T_L\right)\om_\oplus\right]}&=&\dfrac{\sin{\ch}\cos{\th_l}-\cos{\ch}\sin{\th_l}\sin{\ph_l}}{\sin \chM},\nn\\
\sin{\left[\left(T_\oplus-T_L\right)\om_\oplus\right]}&=&-\dfrac{\sin\th_l}{\sin \chM} \cos{\ph_l}.
\label{phM}
\eea
Here, $\ch$ is the colatitude of the experiment, and $\chM$ represents the angle of the magnetic field with Earth's rotation axis. The angles $\th_l$ and $\ph_l$ denote the magnetic field's orientation relative to the local zenith and local East, respectively. Using that for the SMI/LAC experiment $\chM\simeq 49^\circ$, $\th_l\simeq \pi/2$, $\ch=41.3^\circ$ and $\ph_l=84^\circ$, we obtain that 
\beq
\left( T_\oplus-T_L\right)\om_\oplus\simeq -172^\circ.
\eeq 
For the sake of completeness, the relationship between the local sidereal time $T_\oplus$ and the time coordinate $T$ can be expressed as
\cite{dk16},
\beq
T_\oplus\simeq T-\fr{(66.25^\circ- \la)}{360^\circ} 23.934~{\rm hr},
\label{T0}
\eeq
where $\la$ is the longitude of the laboratory in degrees.

Applying the transformation \eqref{LTlinear} to the coefficients in the expression for the frequency shift \eqref{LVfshift} is not straightforward. The nonrelativistic spherical coefficients are specifically designed to have straightforward transformation rules under rotations, as discussed in the preceding subsection. However, their transformations under a boost are notably intricate. Transforming the coefficients necessitates a departure from presenting them as components of spherical tensors to portraying them as components of Cartesian spacetime tensors. This process was thoroughly explained in appendix A and B of \cite{kv18}.

The expression for the frequency shift in the Sun-entered frame at linear-boost order takes the form:
\bea
2\pi \de \nu^{(1)}
&=&
A_0^{(1)}+A_c^{(1)} \cos\om_\oplus T_L+A_s^{(1)} \sin\om_\oplus T_L
\nn\\
&&
+A_C^{(1)} \cos\Om_\oplus T+A_S^{(1)} \sin\Om_\oplus T
\nn\\
&&
+\cos \om_\oplus T_L \left(A_{cC}^{(1)} \cos\Om_\oplus T+A_{cS}^{1)} \sin\Om_\oplus T\right)
\nn\\
&&
+\sin \om_\oplus T_L \left(A_{sC}^{(1)} \cos\Om_\oplus T+A_{sS}^{(1)} \sin\Om_\oplus T\right)
\nn\\
&&
+\cos 2\om_\oplus T_L \left(A_{c2C}^{(1)} \cos \Om_\oplus T+A_{c2S}^{(1)} \sin \Om_\oplus T\right)
\nn\\
&&
+\sin 2\om_\oplus T_L \left(A_{s2C}^{(1)} \cos \Om_\oplus T+A_{s2S}^{(1)} \sin \Om_\oplus T\right)
\nn\\
&&
+A_{c2}^{(1)} \cos 2\om_\oplus T_L+A_{s2}^{(1)} \sin 2\om_\oplus T_L 
\nn\\
&&
+A_{c3}^{(1)} \cos 3\om_\oplus T_L+A_{s3}^{(1)} \sin 3\om_\oplus T_L.
\label{sunshift1}
\eea

The coefficients $A_\xi^{(1)}$ (with the superscript $(1)$) differ from the coefficients $A_\xi^{(0)}$ (with the superscript $(0)$) presented in the previous subsection. The latter represents the amplitudes for the frequency shift $\delta \nu^{(0)}$ at the zeroth-boost order, while the former represents the amplitudes for the frequency shift $\delta \nu^{(1)}$ at the linear-boost order. 

As previously demonstrated \cite{kv15}, the constraints on the Lorentz-violating frequency shift at zeroth-boost order will restrict a different set of SME coefficients than the ones at linear order if the unperturbed states are parity states, as is the case for deuterium.  Consequently, the primary motivation for extending the analysis beyond the zeroth-boost order approximation is to gain access to more SME coefficients. 

Finally, the frequency shift in the Sun-centered frame, up to the linear-boost order, is given by
\beq
\delta \nu =\delta \nu^{(0)}+\delta \nu^{(1)},
\label{sunshift}
\eeq
where $\delta \nu^{(0)}$ is provided in \rf{sunshift0} and $\delta \nu^{(1)}$ in \rf{sunshift1}. From these equations, we can observe that the signals for Lorentz violation include a sidereal variation up to the third harmonic of the sidereal frequency, an annual variation with the first harmonic of the annual frequency, and a mixed variation that is the product of harmonics of the  annual and sidereal frequencies.

In this study, we will narrow our scope to corrections solely related to the pure sidereal variation by disregarding all the terms that contain an annual variation in the expression \rf{sunshift1}. Upon omitting these terms, the frequency shift is
\bea
2\pi \de \nu^{(1)}
&=&
A_0^{(1)}+A_c^{(1)} \cos\om_\oplus T_L+A_s^{(1)} \sin\om_\oplus T_L
\nn\\
&&
+A_{c2}^{(1)} \cos 2\om_\oplus T_L+A_{s2}^{(1)} \sin 2\om_\oplus T_L 
\nn\\
&&
+A_{c3}^{(1)} \cos 3\om_\oplus T_L+A_{s3}^{(1)} \sin 3\om_\oplus T_L.
\label{sunshift1s}
\eea
The main consequence of excluding terms with annual variation is to limit the number of SME coefficients that can be constrained through a sidereal-variation study based on the model presented in this work. Ideally, a time-variation study of the transition frequency should consider these annual-variation-dependent terms. However, in practice, it is more feasible to conduct a sidereal-variation study than to obtain a dataset for the more intricate time-variation signal described by \eqref{sunshift1}.

The amplitudes in the frequency shift \eqref{sunshift1s} are more involved than those presented in the previous subsection. To enhance clarity, the expressions for the amplitudes are presented in three layers.  We can start by noting that the amplitudes take the form
\bea
A_{\xi}^{(1)}  &=&-\sum_{w={\rm p,n}} \sum_{k,d} m^{d-k-3}_\f\,\fr{2}{25-20 m_F^2}\vev{\pmag^k}_{0E}G_{\xi,\f,k}^{(d)0E} \nn \\
&&-\sum_{w={\rm p,n}} \sum_{k,d} m^{d-k-3}_\f\,\fr{2 m'_F-m_F}{9}\vev{\pmag^k}_{0B}G_{\xi, w, k}^{(d)0B}\nn\\
&&+\sum_{w={\rm p,n}} \sum_{k,d} m^{d-k-3}_\f\,\fr{2 m'_F-m_F}{9}\vev{\pmag^k}_{1B}G_{\xi,\f,k}^{(d)1B}\nn\\
&&+2\sum_{k,d} m^{d-k-3}_e\fr{ m'_F+m_F}{9}\vev{|\pvec_e|^k} G_{\xi, e, k}^{(d)0B}\nn\\
&&+4\sum_{k,d} m^{d-k-3}_e\,\fr{ m'_F+m_F}{9}\vev{|\pvec_e|^k}G_{\xi,e,k}^{(d)1B}.
\label{lv1A}
\eea
Here, the values for the momentum-dependent terms are provided in Table \ref{npexp} and Eq.~\eqref{pmage}. The angular momentum quantum numbers correspond to those introduced in \eqref{LVfshift}. The mass of the nucleon is denoted as $m_\f$, and $m_e$ for the electron. The summation over $k$ is restricted to the values $0$, $2$, and $4$. The index $\xi$ identifies the corresponding amplitude in \eqref{sunshift1}, while the index $d$ represents the mass-dimension of the Lorentz-violating operator contributing to the frequency shift.

Following the convention from previous publications \cite{kv15, kv18}, we will only consider the contribution at linear-boost order from Lorentz-violating operators with mass dimensions $3\leq d \leq 8$. This limitation does not alter the general form of the frequency shift given by \eqref{LVfshift}; it only restricts the coefficients for Lorentz violation that our model can constrain.
\renewcommand\arraystretch{2}
\begin{table*}
\caption{
Expressions for the coefficients $G_{0,\f,k}^{(d)0E}$, $G_{0,\f,k}^{(d)0B}$, and $G_{0,\f,k}^{(d)1B}$ in Eq. \eqref{lv1A}}
\setlength{\tabcolsep}{6pt}
\begin{tabular}{ll}
\hline
\hline
Coefficient &  Expression \\
\hline
$G_{0,\f,k}^{(d)0E} $    &  $\fr{1}{4} r_\oplus \om_\oplus s_\ch s_{2\ch}\left[ c_\ph \left(2{V_w}^{(d)XYZ}_{k2}+{V_w}^{(d)XZY}_{k2}+{V_w}^{(d)YZX}_{k2}\right)+2 s_\ph \left({V_w}^{(d)X(XZ)}_{k2}-{V_w}^{(d)Y(YZ)}_{k2}\right)\right]$\\
$G_{c,\f,k}^{(d)0E} $     &  $ r_\oplus \om_\oplus s_\ch c_{\phi }\left[{V_w}^{(d)YYZ}_{k2}c_{\chi }^2+\frac{1}{4} \left({V_w}^{(d)XXY}_{k2}+4 {V_w}^{(d)XYX}_{k2}+{V_w}^{(d)YYY}_{k2}\right) s_{\chi }^2+{V_w}^{(d)Y}_{k2}\right]$\\
                                    &  $+ r_\oplus \om_\oplus s_\ch  s_{\phi } \left[{V_w}^{(d)XZZ}_{k2} c_{\chi }^2+\frac{1}{4} \left(3 {V_w}^{(d)XXX}_{k2}-{V_w}^{(d)YYX}_{k2}\right) s_{\chi }^2+{V_w}^{(d)X}_{k2}\right]$\\
$G_{s,\f,k}^{(d)0E} $     &  $ r_\oplus \om_\oplus s_\ch c_{\phi } \left[{V_w}^{(d)XZZ}_{k2} c_{\chi }^2+\frac{1}{4} \left({V_w}^{(d)YYX}_{k2}+4  {V_w}^{(d)XYY}_{k2}+{V_w}^{(d)XXX}_{k2}\right)s_{\chi }^2+{V_w}^{(d)X}_{k2}\right]$\\                         
                                     & $-  r_\oplus \om_\oplus s_\ch  s_{\phi } \left[{V_w}^{(d)YZZ}_{k2} c_{\chi }^2+\frac{1}{4} \left(3 {V_w}^{(d)YYY}_{k2}-{V_w}^{(d)XXY}_{k2}\right) s_{\chi }^2+{V_w}^{(d)Y}_{k2}\right]$\\
$G_{c2,\f,k}^{(d)0E} $   & $\frac{1}{4} r_\oplus \om_\oplus s_\ch  s_{2 \chi }\left[\left({V_w}^{(d)YZX}_{k2}-{V_w}^{(d)XZY}_{k2}\right) c_{\phi }+2\left({V_w}^{(d)X(XZ)}_{k2}+{V_w}^{(d)Y(YZ)}_{k2}\right)  s_{\phi } \right]$\\
$G_{s2,\f,k}^{(d)0E} $   & $\frac{1}{4} r_\oplus \om_\oplus s_\ch   s_{2 \chi }\left[2\left({V_w}^{(d)X(XZ)}_{k2}+{V_w}^{(d)Y(YZ)}_{k2}\right)  c_{\phi }-\left({V_w}^{(d)YZX}_{k2}-{V_w}^{(d)XZY}_{k2}\right) s_{\phi } \right]$\\
$G_{c3,\f,k}^{(d)0E} $   & $-\frac{1}{4} r_\oplus \om_\oplus  s_{\chi }^3\left[\left({V_w}^{(d)XXY}_{k2}+{V_w}^{(d)YYY}_{k2}\right)  c_{\phi }-\left({V_w}^{(d)XXX}_{k2}+{V_w}^{(d)YYX}_{k2}\right) s_{\phi } \right]$\\
$G_{s3,\f,k}^{(d)0E} $   & $\frac{1}{4} r_\oplus \om_\oplus   s_{\chi }^3\left[\left({V_w}^{(d)XXX}_{k2}+{V_w}^{(d)YYX}_{k2}\right) c_{\phi }+\left({V_w}^{(d)XXY}_{k2}+{V_w}^{(d)YYY}_{k2}\right)  s_{\phi } \right]$\\
$G_{0,\f,k}^{(d)qB} $    &  $\frac{1}{2} r_\oplus \om_\oplus s_\ch ^2\left[2{T_w}^{(d)(XY)}_{qB, k1} c_{\phi }+ \left({T_w}^{(d)XX}_{qB, k1}-{T_w}^{(d)YY}_{qB, k1}\right) s_{\phi } \right] $\\
$G_{c,\f,k}^{(d)qB} $     &$\frac{1}{2} r_\oplus \om_\oplus s_{2\ch} \left({T_w}^{(d)YZ}_{qB, k1} c_{\phi }+{T_w}^{(d)XZ}_{qB, k1} s_{\phi }\right) $\\
$G_{s,\f,k}^{(d)qB} $     & $ \frac{1}{2} r_\oplus \om_\oplus s_{2\ch}\left({T_w}^{(d)XZ}_{qB, k1} c_{\phi }-{T_w}^{(d)YZ}_{qB, k1} s_{\phi }\right) $\\                         
$G_{c2,\f,k}^{(d)qB} $   & $ r_\oplus \om_\oplus s_\ch^2 \left[{T_w}^{(d)[YX]}_{qB, k1} c_{\phi }+\frac{1}{2}\left({T_w}^{(d)XX}_{qB, k1}+{T_w}^{(d)YY}_{qB, k1}\right) s_{\phi }\right] $\\
$G_{s2,\f,k}^{(d)qB} $   &  $ r_\oplus \om_\oplus s_\ch^2\left[\frac{1}{2}\left({T_w}^{(d)XX}_{qB, k1}+{T_w}^{(d)YY}_{qB, k1}\right)c_{\phi }+{T_w}^{(d)[XY]}_{qB, k1}  s_{\phi }\right] $ \\
\hline\hline
\end{tabular}
\label{lv2A}
\end{table*}

The coefficients $G_{\xi, w, k}^{(d)0E}$ and $G_{\xi, w, k}^{(d)qB}$, where $q$ is either $0$ or $1$, depends on the particle $w$, the momentum power $k$ and mass-dimension $d$. The expressions for these coefficients corresponding to the amplitude $A_\xi^{(1)}$ specified by $\xi$ are provided in Table \ref{lv2A}. The first column of the table displays the coefficients, while the second column contains their expressions. For simplicity, the sine and cosine functions in the table are abbreviated as $s_x=\sin{x}$ and $c_x=\cos{x}$.

In the expressions within the table, $\ch$, $\om_\oplus$, and $r_\oplus$ represent the colatitude of the experiment, the sidereal frequency, and Earth’s radius, respectively. The angle $\phi$ is the phase shift between our two definitions of local sidereal time, $T_L$ and $T_\oplus$. It is defined as
\beq
\phi=\om_\oplus (T_\oplus -T_L)
\eeq
and given by \rf{phM}.
The angle takes the values $\ph\simeq -172^\circ$ for the case of the SMI/LAC experiment.
\renewcommand\arraystretch{1.6}
\begin{table}
\caption{Expressions for ${V_w}_{k2}^{(d)J}$ and ${V_w}_{k2}^{(d)JJ_1 J_2}$ in terms of the effective SME coefficients with  $5 \leq d \leq 8$.} 
\setlength{\tabcolsep}{5pt}
\begin{tabular}{ll}
\hline\hline
$\Va{kj}{d}{JK\ldots M}$ & Combination \\
\hline
${V_w}_{22}^{(5)J}$ 	           & 	 $2{a_w}^{(5)TTJ}_\eff+{a_w}^{(5)KKJ}_\eff$	\\
${V_w}_{22}^{(6)J}$ 	           & 	 $-4{c_w}^{(6)TTTJ}_\eff-4{c_w}^{(6)TKKJ}_\eff$	\\
${V_w}_{22}^{(7)J}$	           & 	 $\frac{10}{3}(2{a_w}^{(7)TTTTJ}_\eff-3{a_w}^{(7)TTKKJ}_\eff)$	\\
${V_w}_{22}^{(8)J}$ 	           & 	 $-10{c_w}^{(8)TTTTTJ}_\eff-20{c_w}^{(8)TTTKKJ}_\eff$	\\
${V_w}_{42}^{(7)J}$	           & 	 $\frac{10}{7}({a_w}^{(7)LLKKJ}_\eff+4 {a_w}^{(7)TTKKJ}_\eff)$	\\
${V_w}_{42}^{(8)J}$ 	           & 	 $-\frac{60}{7}( {c_w}^{(8)TLLKKJ}_\eff+2 {c_w}^{(8)TTTKKJ}_\eff)$	\\
${V_w}_{22}^{(5)JJ_1J_2}$ 	 & 	 $-3 {a_w}^{(5)J\Jj{1}\Jj{2}}_\eff-6\de^{J\Jj{1}}{a_w}^{(5)TT\Jj{2}}_\eff$	\\
${V_w}_{22}^{(6)JJ_1J_2}$	 & 	 $12{c_w}^{(6)TJ\Jj{1}\Jj{2}}_\eff+12\de^{J\Jj{1}}{c_w}^{(6)TTT\Jj{2}}_\eff$	\\
${V_w}_{22}^{(7)JJ_1J_2}$	 & 	 $-30{a_w}^{(7)TTJ\Jj{1}\Jj{2}}_\eff-20\de^{J\Jj{1}}{a_w}^{(7)TTTT\Jj{2}}_\eff$	\\
${V_w}_{22}^{(8)JJ_1J_2}$	 & 	 $60{c_w}^{(8)TTTJ\Jj{1}\Jj{2}}_\eff+30\de^{J\Jj{1}}{c_w}^{(8)TTTTT\Jj{2}}_\eff$	\\
${V_w}_{42}^{(7)JJ_1J_2}$          & 	 $-\frac{60}{7}({a_w}^{(7)TTJ\Jj{1}\Jj{2}}_\eff+\de^{J\Jj{1}}{a_w}^{(7)TTLL\Jj{2}}_\eff)$	\\
                              	          & 	 $\hskip 30pt -\frac{30}{7}{a_w}^{(7)LLJ\Jj{1}\Jj{2}}$	\\
${V_w}_{42}^{(8)JJ_1J_2}$	 & 	 $\frac{180}{7}({c_w}^{(8)TTTJ \Jj{1}\Jj{2}}_\eff+{c_w}^{(8)TKKJ \Jj{1}\Jj{2}}_\eff)$	\\
                                        	 & 	 $\hskip 30pt +\frac{180}{7}\de^{J \Jj{1}}{c_w}^{(8)TTTKK\Jj{2}}_\eff$	\\
\hline\hline
\end{tabular}
\label{appb-2}
\end{table}
\renewcommand\arraystretch{1.6}
\begin{table}
\caption{Expressions for ${T_w}^{(d)JJ_1}_{0B, k1}$ in terms of the effective SME coefficients with $3 \leq d \leq 8$. } 
\setlength{\tabcolsep}{2pt}
\begin{tabular}{ll}
\hline\hline
${T_w}^{(d)JJ_1}_{0B, k1}$ & Combination \\
\hline
${T_w}^{(3)JJ_1}_{0B, 01}$	 & 	$-\Htf{w}{3}{\Jj{1}J}$\\
${T_w}^{(4)JJ_1}_{0B, 01}$	 & 	$2\gtf{w}{4}{\Jj{1}(JT)}$\\	                                 
${T_w}^{(5)JJ_1}_{0B, 01}$	 & 	$-3\Htf{w}{5}{\Jj{1}(JTT)}$\\		 
${T_w}^{(6)JJ_1}_{0B, 01}$	 & 	$4\gtf{w}{6}{\Jj{1}(JTTT)}$\\	
${T_w}^{(7)JJ_1}_{0B, 01}$	 & 	$-5\Htf{w}{7}{\Jj{1}(JTTTT)}$\\	                                 	 
${T_w}^{(8)JJ_1}_{0B, 01}$	 & 	$6\gtf{w}{8}{\Jj{1}(JTTTTT)}$	\\		 
${T_w}^{(5)JJ_1}_{0B, 21}$	 & 	$\frac{1}{5}(3\Htf{w}{5}{J(\Jj{1}KK)}+4\Htf{w}{5}{T(\Jj{1}J)T}$\\	
	                                        & 	$\hskip15pt +2\Htf{w}{5}{TKTK}\de_{J\Jj{1}})$\\                                 	 
${T_w}^{(6)JJ_1}_{0B, 21}$	 & 	$-\frac{3}{5}(3\gtf{w}{6}{J(\Jj{1}KK)T}+4\gtf{w}{6}{T(\Jj{1}J)TT}$\\	 
	                                        & 	$+3\gtf{w}{6}{T(\Jj{1}KK)J}+2\gtf{w}{6}{TKTTK}\de_{J\Jj{1}})$\\
${T_w}^{(7)JJ_1}_{0B, 21}$	 & 	$\frac{6}{5}(3\Htf{w}{7}{J(\Jj{1}KK)TT}+6\Htf{w}{7}{T(\Jj{1}KK)TJ}$\\                                 	 
	                                        & 	$+4\Htf{w}{7}{T(\Jj{1}J)TTT}+2\Htf{w}{7}{TKTTTK}\de_{J\Jj{1}})$\\		
${T_w}^{(8)JJ_1}_{0B, 21}$	 & 	$-2(3\gtf{w}{8}{J(\Jj{1}KK)TTT}+9\gtf{w}{8}{T(\Jj{1}KK)TTJ}$\\
	                                        & 	$+4\gtf{w}{8}{T(\Jj{1}J)TTTT}+2\gtf{w}{8}{TKTTTTK}\de_{J\Jj{1}})$\\	                                 	 
${T_w}^{(7)JJ_1}_{0B, 41}$	 & 	$\frac{3}{35}(16\Htf{w}{7}{T(\Jj{1}JKK)T}+5\Htf{w}{7}{J(\Jj{1}KKLL)}$\\		 
	                                        & 	$\hskip15pt +4\Htf{w}{7}{TKKLLT}\de_{J\Jj{1}})$\\	
${T_w}^{(8)JJ_1}_{0B, 41}$	 & 	$-\frac{3}{7}(16\gtf{w}{8}{T(\Jj{1}JKK)TT}+5\gtf{w}{8}{J(\Jj{1}KKLL)T}$\\	                                 	
	                                        & 	$\hskip15pt+5\gtf{w}{8}{T(\Jj{1}KKLL)J}+4\gtf{w}{8}{TKKLLTT}\de_{J\Jj{1}})$\\	 
\hline\hline
\end{tabular}
\label{appb-3}
\end{table}

\renewcommand\arraystretch{1.6}
\begin{table}
\caption{Expressions for ${T_w}^{(d)JJ_1}_{1B, k1}$in terms of the effective SME coefficients with  $5 \leq d \leq 8$. } 
\setlength{\tabcolsep}{2pt}
\begin{tabular}{ll}
\hline\hline
${T_w}^{(d)JJ_1}_{1B, k1}$ & Combination \\
\hline
${T_w}^{(5)JJ_1}_{1B, 21}$ 	 & 	 $-\frac{1}{5}(15\Htf{w}{5}{\Jj{1}(JTT)}+2\Htf{w}{5}{T(\Jj{1}J)T}$\\
                                                 & 	  $\hskip15pt+6\Htf{w}{5}{\Jj{1}(JKK)}-2\Htf{w}{5}{K(\Jj{1}J)K}$	\\
                                                 & 	 $\hskip30pt +6\Htf{w}{5}{TKTK}\de_{J\Jj{1}})$	\\
${T_w}^{(6)JJ_1}_{1B, 21}$        & 	 $\frac{1}{10}(60\gtf{w}{6}{\Jj{1}(JTTT)}+12\gtf{w}{6}{T(\Jj{1}J)TT}$\\
                                                 & 	  $\hskip15pt+48\gtf{w}{6}{\Jj{1}(TJKK)}-18\gtf{w}{6}{K(\Jj{1}JT)K}$\\
                                                 & 	 $\hskip30pt +21\gtf{w}{6}{TKTTK}\de_{J\Jj{1}})$\\
${T_w}^{(7)JJ_1}_{1B, 21}$        & 	 $-\frac{2}{5}(25\Htf{w}{7}{\Jj{1}(TTTTJ)}+6\Htf{w}{7}{T(\Jj{1}J)TTT}$\\
                                                 & 	  $\hskip15pt+30\Htf{w}{7}{\Jj{1}(TTJKK)}-12\Htf{w}{7}{K(TTJ\Jj{1})K}$\\
                                                 & 	 $\hskip30pt +8\Htf{w}{7}{TKTTTK}\de_{J\Jj{1}})$\\
${T_w}^{(8)JJ_1}_{1B, 21}$  	 & 	 $\frac{1}{2}(30\gtf{w}{8}{\Jj{1}(JTTTTT)}+8\gtf{w}{8}{T(\Jj{1}J)TTTT}$\\
                                                  & 	  $\hskip15pt+48\gtf{w}{8}{\Jj{1}(TTTJKK)}-20\gtf{w}{8}{K(\Jj{1}JTTT)K}$\\
                                                  & 	 $\hskip30pt +9\gtf{w}{8}{TKTTTTK}\de_{J\Jj{1}})$	\\
${T_w}^{(7)JJ_1}_{1B, 41}$  	 & 	 $-\frac{3}{35}(70\Htf{w}{7}{\Jj{1}(JTTKK)}-4\Htf{w}{7}{K(J\Jj{1})KLL}$\\
                                                 & 	  $\hskip15pt+15\Htf{w}{7}{\Jj{1}(JLLKK)}+8\Htf{w}{7}{T(J\Jj{1}LL)T}$	\\
                                                 & 	 $\hskip30pt +16\Htf{w}{7}{TKKLLT}\de^{J\Jj{1}})$\\
${T_w}^{(8)JJ_1}_{1B, 41}$ 	 & 	 $\frac{3}{7}(18\gtf{w}{8}{\Jj{1}(JTKKLL)}-2\gtf{w}{8}{K(J\Jj{1}T)KLL}$\\
                                                 & 	  $\hskip15pt+8\gtf{w}{8}{T(KK\Jj{1}J)TT}+42\gtf{w}{8}{\Jj{1}(JLLTTT)}$\\
                                                 & 	 $\hskip30pt +9\gtf{w}{8}{TKKLLTT}\de^{J\Jj{1}})$\\
\hline\hline
\end{tabular}
\label{appb-4}
\end{table}

The quantities ${V_w}_{k2}^{(d)J}$, ${V_w}_{k2}^{(d)JJ_1J_2}$, ${T_w}_{0B,k1}^{(d)JJ_1}$, and ${T_w}_{1B,k1}^{(d)JJ_1}$, appearing in Table \ref{lv2A}, represent specific linear combinations of effective Cartesian coefficients defined in Appendix B of \cite{kv18}. Here, the indices $J$, $J_1$, and $J_2$ correspond to spatial components $X$, $Y$, and $Z$ within the Sun-centered frame.

To simplify the expressions in the table, we used that components enclosed by brackets indicate antisymmetrization, as in
\beq
T^{Z[XY]}=\fr{1}{2!} (T^{ZXY}-T^{ZYX}), 
\eeq
while those enclosed by parentheses imply symmetrization, as in
\beq
 T^{Z(XY)}=\fr{1}{2!}  (T^{ZXY}+T^{ZYX}).
\eeq

All the relevant linear combinations of effective SME coefficients for this work can be found in the aforementioned appendix and are replicated in tables \ref{appb-2},  \ref{appb-3}, and \ref{appb-4} for convenience. The only relevant coefficients missing from the tables are the ${T_w}^{(d)J}_{1B, 01}$ coefficients that are equal to the ${T_w}^{(d)J}_{0B, 01}$ in the Table \ref{appb-3} . The effective SME coefficients, marked with a subscript $\text{eff}$, differ from the conventional Cartesian SME coefficients. They are the smallest linear combination of SME coefficients that effectively can be observed in experiments involving the propagation of free fermions in the presence of Lorentz violation and  they are thoroughly explained in \cite{km13}.  As the usual SME Cartesian coefficients, they transform as Lorentz tensors under observer transformations. Another relevant property of these coefficients is that the $c$-type and $a$-type coefficients are entirely symmetric tensors. Meanwhile, the $\Ht$ and $\gt$ coefficients exhibit antisymmetry in the first two indices and symmetry upon exchange between the others. 

In the expressions for the linear combinations in the tables, the symbol $\delta^{LK}$ denotes the Kronecker delta, which equals 1 when $K=L$ and 0 otherwise. Any repeated Latin index, such as $K$ or $L$, implies a summation over the spatial components $X$, $Y$, and $Z$. Lastly, components enclosed by parentheses imply that they are symmetrized, as previously mentioned.

Finally, the reason for including electron coefficients at linear-boost order, after being disregarded at zeroth-boost order, is because the model tested in the hydrogen-maser experiment \cite{maser} was limited to zeroth-boost order and predicted a sidereal variation with the first harmonic of the sidereal frequency—the signal investigated in the experiment. As discussed in \cite{kv15}, signals that emerged at linear-boost order in hydrogen's experiments, such as sidereal variation with the second harmonic of the sidereal frequency, remain unexplored. Even if hydrogen is potentially more sensitive to the electron coefficients that contribute to $G_{\xi,e,k}^{(d)qB}$ in \eqref{lv1A}, they remain unconstrained, and we cannot disregard them as we did at zeroth-boost order.  

\section{Prospects for the SMI/LAC experiment}
\label{secprosp}

This section investigates the potential of using deuterium ground-state transitions $F=1/2\rightarrow F=3/2$, with $\Delta m_F=0$, to detect signals for Lorentz violation. To differentiate between the two transition frequencies, we designate the one with $m_F=1/2$ as $\nu_{\si_1}$ and the other with $m_F=-1/2$ as $\nu_{\si_2}$. From the results of the preceding section applied to the case of $m_F=1/2$ and $m_F=-1/2$ with $m'_F=m_F$, we can derive the respective frequency shifts, $\delta \nu_{\si_1}$ and $\delta \nu_{\si_2}$, in the Sun-centered frame.

Within the context of the perturbation \rf{nr}, the $\mathcal{V}$-type nonrelativistic coefficients are referred to as the spin-independent coefficients, while the $\T$-type coefficients are regarded as the spin-dependent ones. This distinction arises from associating the latter with terms containing the Pauli matrices in \rf{nr}. From \rf{LVfshift} with $m’_F=m_F$, we can observe that the contributions of the spin-independent coefficients to the frequency shift are even functions of $m_F$, while the spin-dependent ones are odd functions. Given that the only distinction between the transition frequencies $\nu_{\si_1}$ and $\nu_{\si_2}$ is the sign of $m_F$, the difference 
\beq
\nu_{\si 1-\si 2}=\nu_{\si_1} -\nu_{\si_2}
\eeq
is solely sensitive to the spin-dependent coefficients, while 
\beq
\nu_{\si_1+\si_2}=\nu_{\si_1}+\nu_{\si_2}
\eeq
is sensitive to the spin-independent ones.

At zeroth-boost order, the expected signal for Lorentz violation manifests as a first-harmonic sidereal variation of the frequency difference $\nu_{\si_1-\si_2}$. Meanwhile, the variation in $\nu_{\si_1+\si_2}$ contains contributions up to the second harmonic of the sidereal frequency. The reason why $\nu_{\si_1-\si_2}$ contains only contributions from the first harmonic of $\om_\oplus$ is that the amplitudes associated with the second harmonic of $\om_\oplus$ depend solely on the spin-independent coefficients, as depicted in \rf{A2}. 

We will assess the potential of the SMI/LAC experiment for testing Lorentz and CPT symmetry by assuming that the experiment reaches a feasible sensitivity of 10 Hz on any of the amplitudes for the sidereal variation of the frequencies $\nu_{\si_1-\si_2}$ and $\nu_{\si_1+\si_2}$ \cite{SMIpriv}. For now, we will limit our focus to the zeroth-boost order terms with Table \ref{sen} containing our estimates for the SMI/LAC-experiment sensitivity to the nonrelativistic coefficients. The coefficients listed represent proton and neutron coefficients, as our model does not distinguish between the contributions from both. Additionally, the limits are on the coefficients associated with CPT-odd or CPT-even operators as introduced in the decomposition presented in \rf{cpt}.

\begin{table}
\caption{Potential sensitivity of the SMI/LAC experiment on the nonrelativistic coefficients.}
\renewcommand\arraystretch{1.6}
\setlength{\tabcolsep}{10pt}
\begin{tabular}{cl}
\hline\hline
Coefficient &  Constraint on\\[-6pt]
$\K$ & $|\Re{\K}|, |\Im{\K}|$
\\\hline
$\sHzBnrf{\f}{011},\,\sgzBnrf{\f}{011}$ & $<4\times 10^{-22}$ GeV\\
$\sHoBnrf{\f}{011},\,\sgoBnrf{\f}{011}$ & $<2\times 10^{-22}$ GeV\\
$\sHzBnrf{\f}{211},\,\sgzBnrf{\f}{211}$ & $<3\times 10^{-20}$ GeV$^{-1}$\\
$\sHoBnrf{\f}{211},\,\sgoBnrf{\f}{211}$ & $<6\times 10^{-20}$ GeV$^{-1}$\\
$\sHzBnrf{\f}{411},\,\sgzBnrf{\f}{411}$ & $<7\times 10^{-20}$ GeV$^{-3}$\\
$\sHoBnrf{\f}{411},\,\sgoBnrf{\f}{411}$ & $<2\times 10^{-19}$ GeV$^{-3}$\\
$\scnrf{\f}{221},\,\sanrf{\f}{221}$ & $<4\times 10^{-20}$ GeV$^{-1}$\\
$\scnrf{\f}{222},\,\sanrf{\f}{222}$ & $<6\times 10^{-20}$ GeV$^{-1}$\\
$\scnrf{\f}{421},\,\sanrf{\f}{421}$ & $<2\times 10^{-19}$ GeV$^{-3}$\\
$\scnrf{\f}{422},\,\sanrf{\f}{422}$ & $<4\times 10^{-19}$ GeV$^{-3}$\\
\hline\hline
\end{tabular}
\label{sen}
\end{table}

The most stringent constraints on nonrelativistic nucleon coefficients stem from four experiments. These include time-variation studies conducted using a $^{129}$Xe-$^3$He comagnetometer \cite{xema, xema2, Xe09, Xe14}, a $^{21}$Ne-Rb-K comagnetometer \cite{NeRb}, an optical hydrogen transition \cite{ma13}, and the aforementioned hydrogen maser \cite{maser}. The limits derived from the first two systems are in \cite{kv18}, while the rest are in \cite{kv15}. 

Among these experiments, both the hydrogen maser and the $^{129}$Xe-$^3$He comagnetometer experiments are sensitive only to the spin-dependent coefficients. According to the Schmidt model, the Xe-He experiment can exclusively access neutron coefficients. Independently of any model, the hydrogen experiment can solely detect proton ones, besides the electron coefficients, as hydrogen does not contain neutrons.

The limits derived from the Xe-He experiment on the neutron spin-dependent coefficients are contained in Table VI of \cite{kv18}. They are in the order of $10^{-33}{\rm GeV}$ for $k=0$, $10^{-31}{\rm GeV}^{-1}$ for $k=2$, and $10^{-29}{\rm GeV}^{-3}$ for $k=4$. Therefore, they exceed our estimates for the deuterium experiment by around $10$ orders of magnitude.

Table III of \cite{kv15} contains the hydrogen-maser limits on the spin-dependent proton coefficients; they are in the order of $10^{-27}{\rm GeV}$ for $k=0$, $10^{-16}{\rm GeV}^{-1}$ for $k=2$, and $10^{-6}{\rm GeV}^{-3}$ for $k=4$. For the $k=0$ coefficients, the prospects for the deuterium experiment are unfavorable as the current limits from the hydrogen-maser experiment remain approximately $7$ orders of magnitude superior to our estimates. The situation is different for the ones with $k=2$ and $k=4$. The deuterium experiment could improve the limits on the coefficients $k=2$ by $4$ orders of magnitude and to the ones with $k=4$ by 14 orders of magnitude.

The reported constraints from the remaining two experiments exclusively pertain to the spin-independent nonrelativistic coefficients. However, the discussion concerning the annual variation study of the 1S-2S transition in hydrogen \cite{ma13} will be postponed until we move to the prospects for the deuterium experiment at linear-boost order.

The  $^{21}$Ne-Rb-K experiment has access limited to neutron nonrelativistic spin-independent coefficients according to the Schmidt model \cite{kv18}. Constraints on these nonrelativistic coefficients, obtained from this experiment, are in Table X of \cite{kv18}. These limits are in the order of $10^{-29}{\rm GeV}^{-1}$ for $k=2$ and $10^{-27}{\rm GeV}^{-3}$ for $k=4$. Therefore, they exceed our estimates for the SMI/LAC experiment by at least seven orders of magnitude. The proton spin-independent coefficients in Table \ref{sen} are unconstrained. Consequently, the SMI/LAC experiment holds the potential to establish first bounds on these coefficients.

Moving on to the prospects at linear-boost order, the observable signals for Lorentz violation regarding the frequency difference $\nu_{\si_1-\si_2}$ at this order include a sidereal variation with contributions up to the second harmonic of the sidereal frequency. For $\nu_{\si_1+\si_2}$, the observable signals comprise variations extending from the first to the third harmonic of $\omega_\oplus$. The type of SME coefficients that can be constrained by studying these signals are the effective Cartesian coefficients mentioned in the previous section. Existing constraints on these coefficients are from the Xe-He experiment and an annual variation study of the 1S-2S transition with hydrogen.

For the sake of the discussion, we can estimate the sensitivity of the SMI/LAC experiment to the effective Cartesian coefficient $\K^{(d)\mu_1\ldots\mu_{d-2}}_\eff$ to be in the order of $10^{-14}\, {\rm GeV}^{4-d}$ or less. Comparing this value to the limits on the neutron $g$- and $H$-type effective coefficients obtained from the Xe-He experiment, in Table IX of \cite{kv18}, we conclude they are superior to our estimates for the SMI/LAC experiment by more than $10$ orders of magnitude.

The prospects for the proton $a$- and $c$-type coefficients are promising. The constraints on the proton coefficients obtained from the hydrogen experiment, see Table of \cite{kv15}, could be improved from 6 to 16 orders of magnitude by the SMI/LAC experiment. Finally, all the neutron spin-independent and the proton spin-dependent effective coefficients accessible by the deuterium experiment remain unconstrained, presenting the SMI/LAC experiment with an opportunity to establish first bounds on these coefficients.

As previously mentioned, the electron coefficients contributing at linear-boost order are unconstrained, presenting an opportunity for the SMI/LAC experiment to establish first bounds on these coefficients. It is important to note that these electron coefficients exclusively contribute to the frequency combination $\nu_{\si_1-\si_2}$, given that only the electron spin-dependent coefficients contribute to the Lorentz-violating frequency shifts.

It is worth noting that the transitions from $F'=1/2$ to $F=3/2$ with $\Delta m_F=0$, considered in this section, may not necessarily exhibit the highest sensitivity to the SME coefficients. For instance, transitions such as $m_F=-m'_F=1/2$ are thrice as sensitive to the spin-dependent nucleon coefficients as the case where $m'_F=m_F$. Similarly, transitions such as $m_F=3m'_F=3/2$ are twice as sensitive to the electron spin-dependent terms. We must be careful about these statements as they follow the strong assumption that the experimental constraint on the Lorentz-violating frequency shifts for all transitions are the same.

Before concluding this section, it is worthwhile to compare our estimates for the SMI/LAC experiment with a proposed, in \cite{kv18}, reanalysis of the data used in a $^{133}$Cs-fountain-clock experiment \cite{cs06,cs17}. This reanalysis of the Cs experiment could impose constraints that might surpass our estimated SMI/LAC limits on the spin-independent proton coefficients by at least three orders of magnitude based on the potential sensitivities reported in Table V of \cite{kv18}.

Based on the discussion in this section, it is evident that the prospects for the SMI/LAC experiment are promising, as they hold the potential to enhance current limits on nonminimal SME coefficients and establish initial limits on many of them.

\section{Summary}

This study improves a previous model for testing Lorentz and CPT symmetry using time-variation studies of the hyperfine-Zeeman transition within the ground state of deuterium. It begins by obtaining the Lorentz-violating correction to deuterium's spectrum under the influence of a weak magnetic field by using a well-established parametrization of the deuteron ground-state wave function. It is worth noting that this is the first time, within the nonminimal SME context, that a nuclear model beyond the Schmidt model was used to obtain the nucleon contribution to the Lorentz-violating energy shift.

We determined the Lorentz-violating frequency shift for the transitions within the ground state of deuterium in terms of the laboratory frame SME coefficients. Subsequently, we described the transformation from the laboratory to the Sun-centered frame. The Lorentz transformation matrix was expanded as a power series of the laboratory-frame boost velocity within the Sun-centered frame and truncated to linear order. Applying the approximated transformation matrix, we obtained a general expression for the frequency shift at zeroth- and linear-boost orders.

From the expression for the frequency shift in terms of the Sun-centered frame coefficients, this study identified the signals for Lorentz violation revealed as time-variations of the transition frequency induced by the motion of the Earth-based experiment within the Sun-centered frame. When the frequency shift is limited to zeroth-boost order, the signal is a sidereal variation entailing contributions up to the second harmonic of the sidereal frequency, and if we include the linear-boost order term, the third harmonic of the sidereal frequency contributes. This work also identified annual variation and more complex types of variation as signals for Lorentz violation accessible via deuterium ground-state spectroscopy. However, this study provides explicit expressions solely for the sidereal-variation amplitudes.

Using the model developed in this work, we estimated the sensitivity of the ongoing SMI/LAC deuterium experiment to the relevant SME coefficients to compare them to all current best limits on Lorentz violation. We concluded that the SMI/LAC experiment holds considerable potential to establish more stringent bounds than the existing ones on most proton nonrelativistic spin-dependent coefficients and first-time limits on all proton spin-independent ones. It can impose first-time limits on $a$- and $c$-type neutron coefficients and on a $g$- and $H$-type electron ones. 

In summary, the SMI/LAC deuterium experiment can impose first-time or improved limits on many proton, neutron, and electron SME coefficients.  

\section*{Acknowledgments}

We extend our gratitude to Amit Nanda, Martin Simon, and Eberhard Widmann for their invaluable assistance in providing us with detailed insights and a deeper understanding of the SMI/LAC experiment.

\end{document}